\documentstyle[prl,twocolumn,aps,epsf]{revtex}
\def\img{\mbox{i}}
\def\const{\mbox{const.}}
\def\tA{t_{\mbox{A}}}
\def\tB{t_{\mbox{B}}}
\def\VA{V_{\mbox{A}}}
\def\VB{V_{\mbox{B}}}
\def\JA{J_{\mbox{A}}}
\def\JB{J_{\mbox{B}}}
\def\JAA{J_{\mbox{AA}}}
\def\JAB{J_{\mbox{AB}}}
\def\JBA{J_{\mbox{BA}}}
\def\img{\mbox{i}}
\def\H{{\cal H}}
\def\v#1{\mbox{\boldmath$#1$}}

\begin{document}
\draft
\title{Quasiperiodic Hubbard chains}
\author{Kazuo Hida}
\address{Department~of~Physics,
 Faculty of Science,\\ Saitama University, Urawa, Saitama 338-8570, JAPAN}

\date{Received \today}

\maketitle

\begin{abstract}
{
Low energy properties of half-filled Fibonacci Hubbard models are studied by weak coupling renormalization group and density matrix renormalization group method. In the case of diagonal modulation, weak Coulomb repulsion is irrelevant and the system behaves as a free Fibonacci chain, while for strong Coulomb repulsion, the charge sector is a Mott insulator and the spin sector behaves as a uniform Heisenberg antiferromagnetic chain. The off-diagonal modulation always drives the charge sector to a Mott insulator and the spin sector to a Fibonacci antiferromagnetic Heisenberg chain.}

\end{abstract}

\pacs{75.10.Jm, 75.50.Kj, 71.30+h, 71.10-w, 71.23Ft}

Since the discovery of high $T_c$ oxide superconductors and heavy fermion materials, the strongly correlated electron system has been the most important subject of recent condensed matter physics. Even in the insulating phase, the quantum magnetism in these and related materials have been attracting the wide interest from theory and experiment\cite{miyako}. Another remarkable finding in recent condensed matter physics is the discovery of quasicrystals\cite{shecht}. The electronic states in quasicrystals are not trivial even in the simplest case of one dimensional free fermions.  For the Fibonacci lattice, the beautiful multifractal structure of the single particle spectrum and the wave function have been revealed by means of the renormalization group method\cite{kkt1,jh1}. 

Nevertheless the interplay between the quasiperiodicity and strong correlation in quantum magnetism has been rarely studied except for the recent bosonization\cite{vidal} and density  matrix renormalization group (DMRG) studies for one-dimensional Heisenberg chains.\cite{kh_h} Experimentally, several kinds of quasicrystals with local magnetic moments have been synthesized recently\cite{sato1}.  In this respect, the quantum magnetism in quasiperiodic systems must be a promising field in the condensed matter physic of next decade.

Although one dimensional quasiperiodic magnetic material is not available at present, one of the plausible candidate would be an artificial structure such as quantum dot array\cite{dot1}. In such  realistic situations, the iteneracy of electrons becomes important. For the realization of one dimensional Fibonacci antiferromagnet, it is necessary to specify the parameter regime in which quasiperiodicity comes into play. In the present work, we therefore investigate the Fibonacci Hubbard model in which the coupling between spin and charge degrees of freedom produces a rich variety of ground states even in the half-filled case. 

Our Hamiltionan is given by,
\begin{eqnarray}
\label{eq:ham_d}
\H^d &=& \sum_{i=1}^{N-1} -t[a^{\dagger}_{i,\sigma}a_{i+1,\sigma}+a^{\dagger}_{i+1,\sigma}a_{i,\sigma}]  \nonumber \\
&+& \sum_{i=1}^{N} \left[ V_{\alpha_i}(n_{i,\uparrow}+n_{i,\downarrow}-1) \right.\nonumber \\
&+&\left.U(n_{i,\uparrow}-1/2)(n_{i,\downarrow}-1/2)\right],\ \ \ (t, U > 0),
\end{eqnarray}
for diagonal modulation and
\begin{eqnarray}
\label{eq:ham_o}
\H^o &=& \sum_{i=1}^{N-1} -t_{\alpha_i}[a^{\dagger}_{i,\sigma}a_{i+1,\sigma}+a^{\dagger}_{i+1,\sigma}a_{i,\sigma}] \nonumber \\
&+& \sum_{i=1}^{N} U(n_{i,\uparrow}-1/2)(n_{i,\downarrow}-1/2),\ \ \ (t_{\alpha_i}, U > 0),
\end{eqnarray}
for off-diagonal modulation. The superfices $d$ and $o$ represent the diagonal and off-diagonal modulations, respectively. The operators $a^{\dagger}_{i,\sigma}$ and $a_{i,\sigma}$ are creation and anihilation operators of fermions with spin $\sigma (=\uparrow \mbox{or} \downarrow)$ and $n_{i,\sigma}=a^{\dagger}_{i,\sigma}a_{i,\sigma}$. The open boundary condition is assumed. The on-site coulomb interaction is denoted by $U$. The transfer integral $t_{\alpha_i}$'s ($=\tA$ or $\tB$) or the on-site potential $V_{\alpha_i}$'s ($=\VA$ or $\VB$) follow the Fibonacci sequence generated by the substitution rule $A \rightarrow A B, \ B \rightarrow A$. The modulation amplitudes are defined by $\Delta t = \tA-\tB$ and $\Delta V = \VA -\VB$.  For the off-diagonal modulation case, the average transfer integral $t$ is defined by $t \equiv (\tA\phi+\tB)/(1+\phi)$ where $\phi$ is the golden ratio $\phi \equiv (1+\sqrt{5})/2$. In the rest of this paper, we concentrate on the half-filled case.

First, we employ the bosonization method in the weak coupling limit $U, \mid\Delta t\mid,  \mid\Delta V\mid << t$, to obtain the following bosonized Hamiltonian

\begin{equation} \label{Hbos}
\H_B^{d,o}=\H_0+\H_W^{d,o},
\end{equation}
where
\begin{eqnarray} 
\H_0&=&\H_{\rho}+\H_{\sigma} \nonumber \\
\label{H0bos}
\H_{\mu}&=&{1\over 2\pi \alpha}\int dx \left[(u_{\mu} K_{\mu})(\pi \Pi_{\mu})^2+\left({u_{\mu}\over K_{\mu}}\right)
(\partial_x \phi_{\mu})^2\right]\nonumber \\
&+&{y_{\mu}v_F\over 2\pi\alpha^2}\int dx \cos \left[2\sqrt{2}\phi_{\mu}\right],\ (\mu = \rho \mbox{ or } \sigma ) \nonumber  \\
\label{H1bos}
\H_W^d &=& \frac {\Delta V}{\pi \alpha} \int dx \,W(x)  e^{\img\pi x/a}  \cos\left[\sqrt{2}\phi_{\rho}(x)\right]  \cos\left[\sqrt{2}\phi_{\sigma}(x)\right], \nonumber  \\
\label{hquasi_d}
\H_W^o &=& \frac{2\Delta t}{\pi \alpha} \int dx \,W(x)  e^{\img\pi x/a} \sin\left[\sqrt{2}\phi_{\rho}(x)\right]  \cos\left[\sqrt{2}\phi_{\sigma}(x)\right],  \nonumber 
\label{hquasi_o}
\end{eqnarray}
with 
\begin{eqnarray} \label{para}
y_{\mu}&=&Ua/\pi v_F,\ v_F = 2ta \nonumber \\
K_{\sigma}&=&\frac{1}{\sqrt{1-Ua/\pi v_F}},\ K_{\rho}=\frac{1}{\sqrt{1+Ua/\pi v_F}} \nonumber \\
u_{\sigma}&=&v_F\sqrt{1-Ua/\pi v_F},\ u_{\rho}=v_F\sqrt{1+Ua/\pi v_F} \nonumber 
\end{eqnarray}
The boson fields $\phi_{\rho}$ and $\phi_{\sigma}$ represent the charge and spin degrees of freedom, respectively. The momentum densities conjugate to them are denoted by $\Pi_{\rho}$ and $\Pi_{\sigma}$. The lattice constant, fermi velocity are denoted by $a$ and $v_F$. The ultraviolet cut-off denoted by $\alpha$ is of the order of $a$. The function $W(x)$ represents the Fibonacci modulation of amplitude unity.

Following Vidal et al.\cite{vidal}, we obtain  the weak coupling renormalization group (WCRG) equations for the coupling constants by the standard technique\cite{gsf} as,
\begin{eqnarray}
{dK_{\rho}\over dl}&=&-K_{\rho}^2\left(\frac{y_{\rho}^2}{2}+ G(l)\right), \label{recy1}\\
{dK_{\sigma}\over dl}&=&-K_{\sigma}^2\left(\frac{y_{\sigma}^2}{2}+ G(l)\right), \label{recy2}\\
{dy_{\rho}\over dl}&=&(2-2K_{\rho})  y_{\rho} \pm 2G(l), \label{recy3} \\
{dy_{\sigma}\over dl}&=&(2-2K_{\sigma})  y_{\sigma} -2G(l), \label{recy4} \\
{dy_q\over dl}&=&(2-K_{\rho}/2-K_{\sigma}/2-y_{\rho}/2-y_{\sigma}/2)y_q, \label{recy5} \\
G(l) &=& \sum_{\varepsilon=\pm 1}\sum_q y_q^2 R\left[(q + \varepsilon \pi / a) \alpha(l) \right], 
\label{recy6}
\end{eqnarray}
where $y_q(0)=\alpha \lambda \hat W(q) / v_F$ with $\lambda =\Delta V/2$ for the diagonal modulation and  $\lambda =\Delta t$ for the off-diagonal modulation. The Fourier components of $W(x)$ is denoted by $\hat W(q)$ whose explicit form is given in \cite{vidal}. The renormalized short distance cut-off is given by $\alpha(l)=\alpha e^l$.
In eq.(\ref{recy6}), the summation over $q$ is performed for
$q=2\pi m/n$ with $m\in [1,n-1]$ where $n$ is the generation of the Fibonacci sequence and  $R$ is the Gaussian ultraviolet regulator $R(x)=e^{-x^2}$. The $+ (-) $ sign in eq.(\ref{recy3}) is for the off-diagonal (diagonal) modulation case. The corrections to the velocities $u_{\rho}$ and $u_{\sigma}$ are neglected since they give the higher order corrections to the renormalization of other quantities. Similar set of equations for the alternating potential are derived by Tsuchiizu and Suzumura\cite{ts1}. 
It should be noted that the scaling dimension of the quasiperiodic modulation term is non-trivial because of the presence of self-similar function $W(x)$, so that we have to resort to numerical calculation to solve the WCRG equations. 

We have also carried out the numerical calculation using the DMRG method to obtain insight into the strong coupling regime which is inaccessible by the WCRG calculation. In the numerical calculation, we take $t=1$ to fix the energy unit.  To reveal the bulk properties of the Fibonacci chains, it is useful to investigate the behavior of the average of physical quantities over all possible finite length subsequences of infinite Fibonacci chains as discussed in \cite{kh_h,ig1}. It should be also noted that the number of the $n$-membered subsequence is equal to $n+1$\cite{penrose}. The number $m$ of the states kept on each DMRG step ranged from 120 to 300 depending on the values of parameters.  The convergence with respect to $m$ is checked. If weak $m$-dependence remains around $m=300$, the $m$-extrapolation is carried out for each energy eigenvalue $E(m)$ using the extrapolation formula $E(m) \simeq E(\infty) + c/m^2$\cite{narushima}.
\begin{figure}
\epsfxsize=70mm 
\centerline{\epsfbox{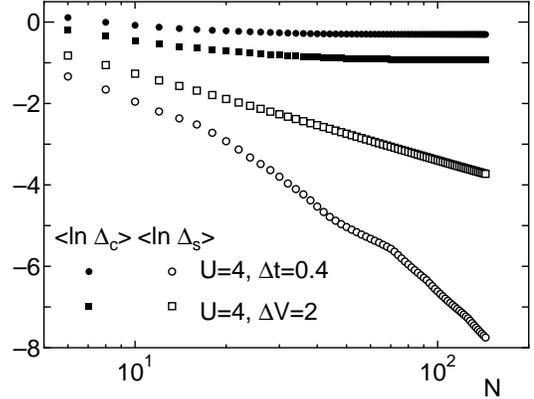}}
\caption{The $N$-dependence of $<\ln\Delta_{s,c}>$  for the Fibonacci Hubbard model with $U=4$ for  diagonal modulation with $\Delta V = 2.0$ (squares) and off-diagonal modulation with $\Delta t = 0.4$ (circles). Filled (open) symbols represent the charge (spin) gap .}
\label{fig1}
\end{figure}
Numerical solutions of the WCRG equations are classified into two categories according to the behavior of $y_{\rho}$. In the following, we discuss these two regimes separately.

\noindent
(a){\it Strong coupling regime}

If $y_{\rho}$ increases in the course of renormalization, it grows more rapidly than other parameters, so that the phase $\phi_{\rho}$ is fixed to $\pi/2\sqrt{2}$. The growth of the Fibonacci modulation term $G$ is suppressed as expected from $-y_{\rho}$ term in the RHS of eq. (\ref{recy5}). In this case, the charge gap opens and the ground state is a Mott insulator.  For diagonal modulation, the Fibonacci modulation term, which is proportional to $\cos \sqrt{2}\phi_{\rho}$, vanishes. Thus the low energy behavior of the spin sector is described as an antiferromagnetic uniform Heisenberg chain. On the other hand, for off-diagonal modulation, the Fibonacci modulation term is proportional to $\sin \sqrt{2}\phi_{\rho}$ which is fixed to 1. Therefore the spin sector is renormalized to the Fibonacci antiferromagnetic Heisenberg chain whose behavior is discussed in detail in \cite{vidal,kh_h}. It should be noted that $K_{\rho}$ is always less than unity  because $K_{\rho}(0) < 1$ and $\frac{dK_{\rho}}{dl} <0$. For the off diagonal modulation, therefore, $\frac{dy_{\rho}}{dl}$ is always positive and the above behavior is always realized.

Typical cases are numerically demonstrated by the DMRG method in Fig. \ref{fig1} with $U=4$ for the off-diagonal modulation with $\Delta t =0.4$ and  diagonal modulation with $\Delta V = 2.0$. The system size dependence of the average of the logarithm of the spin gap $\Delta_s$ and charge gap $\Delta_c$ are shown.  It is verified that the charge gap tends to a finite value as $N \rightarrow \infty$ for both cases. For the diagonal modulation, the spin gap behaves as $<\ln\Delta_s> \simeq -\ln N$ with slope unity which is the Luttinger liquid behavior of the uniform Heisenberg chain.  For the off-diagonal modulation, the size dependence of the spin gap is well fitted by the formula $<\ln\Delta_s> \sim -N^{\omega}$ as in the Fibonacci Heisenberg chain\cite{kh_h}. It should be noted that no trace of Fibonacci modulation remains in the size dependence of the spin gaps of the diagonal case even though the modulation amplitude is 5 times larger than the off-diagonal case.

\begin{figure}
\epsfxsize=90mm 
\centerline{\epsfbox{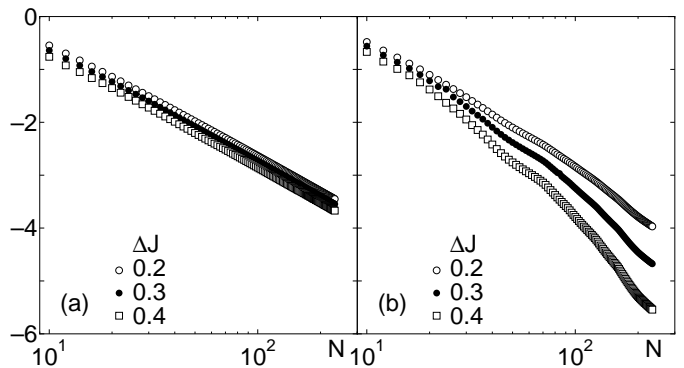}}
\caption{The $N$-dependence of $<\ln\Delta_s>$ for the Heisenberg model with (a) Fibonacci pairwise modulation  and (b) Fibonacci modulation with $J = 1.0$. }
\label{fibhei}
\end{figure}

In the limit of strong $U >> \tA, \tB$, our Hubbard model can be mapped onto the Heisenberg model as,
\begin{equation}
\label{eq:ham_h}
\H_{\mbox{H}} = \sum_{i=1}^{N-1} 2J_{\alpha_i}\v{S}_{i}\v{S}_{i+1},\ \ \ (J_{\alpha_i} > 0),
\end{equation}
where $\v{S}_{i}$'s are the spin 1/2 operators. In the case of off-diagonal modulation the exchange couplings $J_{\alpha_i}$'s ($=\JA$ or $\JB$) follow the Fibonacci sequence as, $\JA=2\tA^2/U$ and $\JB=2\tB^2/U$. On the other hand, for the diagonal modulation, the exchange coupling is determined by $U$ and $V_{\alpha}$ of the sites on the both ends of the bond. Therefore the exchange coupling can be indexed by the pair of letters which appear in Fibonacci sequence as $\JAA = 2t^2/U$ and $\JAB=\JBA= t^2/(U+\VA-\VB)+ t^2/(U+\VB-\VA)$. The sequence BB does not appear. We call this type of modulation as pairwise Fibonacci modulation. The system size dependence of the energy gap calculated by the DMRG method for both types of antiferromagnetic Heisenberg chains are shown in Fig. \ref{fibhei} with $\Delta J =\mid \JA-\JB\mid$ or $\mid \JAA-\JAB\mid$. The energy unit is $J \equiv (\phi\JA+\JB)/(1+\phi)$ or $(\phi\JAA+\JAB)/(1+\phi)$. As expected, the spin excitations scales as $\Delta_s \sim \exp(-cN^{\omega})$ for the Fibonacci Heisenberg chain and as $\Delta_s \sim -\ln N$ for the pairwise Fibonacci chain. 

\noindent
(b) {\it Asymptotically free regime}

\begin{figure}
\epsfxsize=90mm 
\centerline{\epsfbox{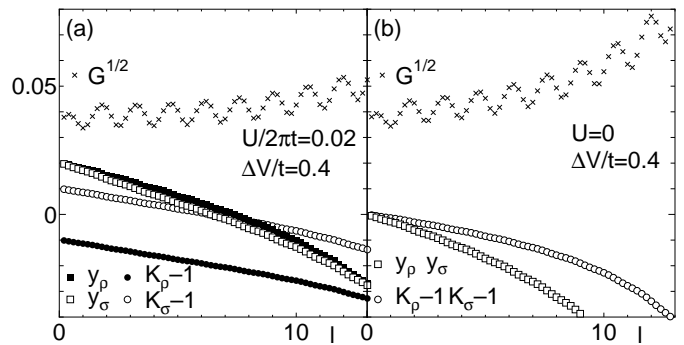}}
\caption{The $l$-dependence of the renormalized parameters with $\Delta V/t = 0.4$ with (a) $U/2\pi t=0.02$ and (b) $U=0$. The ultraviolet cut-off $\alpha$ is taken equal to $a$. }
\label{figfl}
\end{figure}

\begin{figure}
\epsfxsize=90mm 
\centerline{\epsfbox{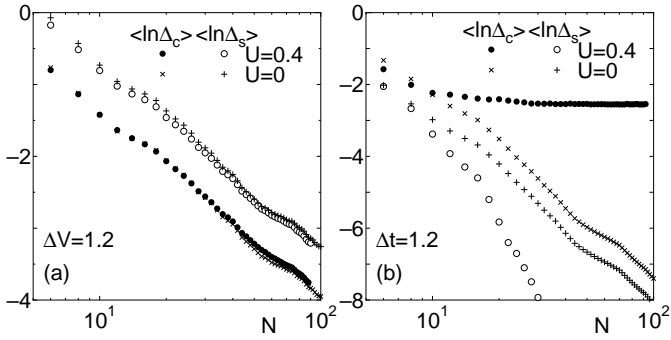}}
\caption{The $N$-dependence of $<\ln\Delta_{s,c}>$  for the Fibonacci Hubbard model with $U=0.4$ and $U=0$ for (a) diagonal modulation with $\Delta V = 1.2$ and (b) off-diagonal modulation with $\Delta t = 1.2$. Filled (open) symbols represent the charge (spin) gap .}
\label{fig2}
\end{figure}

For the diagonal modulation case, the competition between the quasiperiodic modulation and the Coulomb interaction can take place for small $U$. In this case, both $y_{\rho}$ and $y_{\sigma}$ are renormalized to negative values and the Fibonacci modulation term $G$ grows under renormalization, so that none of them are dominating. Typical example of the numerical solution of WCRG equations is shown in Fig. \ref{figfl}(a) for $\Delta V/t = 0.4$ and $U/2\pi t=0.02$. Comparing this behavior with that of the free fermions with Fibonacci potential shown in Fig. \ref{figfl}(b) with $\Delta V/t = 0.4$, we find both flows are quite similar. This implies that the main contribution to the renormalized quantities are generated by the back scattering due to the Fibonacci modulation and the bare Coulomb interaction does not make essential contribution. Therefore we conclude that the weak Coulomb interaction is irrelevant and the low energy spectrum scales as $<\ln \Delta_{s,c}> \sim -z\ln N$ similarly to the free case. The numerical results by DMRG shown in Fig. \ref{fig2}(a) for $U=0.4$ and $\Delta V=1.2$ also support this conclusion if compared with the exact diagonalization results for the free case with $\Delta V=1.2$ shown in the same figure.  For comparison, Fig. \ref{fig2}(b) shows the same quantities for the off-diagonal modulation with $U=0.4$ and $\Delta t=1.2$. In contrast to the diagonal modulation case, there is a clear evidence of the finite charge gap. The spin gap decreases too rapidly to estimate the precise value for large systems $N > 36$. It is, however, consistent with the Fibonacci Heisenberg type behavior $<\ln\Delta_s> \sim -N^{\omega}$\cite{kh_h} rather than the power law $<\ln\Delta_s> \simeq -z\ln N$. 

Following the argument of \cite{jh1}, the low temperature behavior of the thermodynamic quantities are deduced from the above scaling behavior of the low energy spectrum. For the off-diagonal modultion, the magnetic susceptibility $\chi$ behaves as $\chi \sim 1/(T(\ln T)^{1/\omega})$ and magnetic specific heat $C$ as $C \sim 1/(\ln T)^{1+1/\omega}$. This also holds for the Fibonacci Heisenberg chains. For the diagonal modulation with large $U$, we expect the Luttinger liquid behavior $\chi \sim \const$ and $C \sim T$, while for the diagonal modulation with small $U$, we expect the free Fibonacci chain behavior $\chi \sim 1/T^{1/z-1}$ and $C \sim 1/T^{1/z}$.

In summary, we find that the Fibonacci repulsive Hubbard model at half-filling shows a variety of ground states depending on the types and strength of modulation in contrast with the free fermion Fibonacci chains which is critical irrespective of the type or strength of modulation. For the off-diagonal modulation, the effect of Coulomb interaction is most drastic. The ground state is always a Mott insulator and the spin sector behaves as an antiferromagnetic Fibonacci Heisenberg chain. On the contrary, for the diagonal modulation, both spin and charge sectors behave as free Fibonacci chains if the Coulomb interaction is weak enough. This implies that the conventional free theories\cite{kkt1} are approximately applicable to the diagonal modulation case as far as the electron-electron interaction is weak. This is in contrast to the case of spinless fermion chains in which nearest neighbour repulsive Coulomb interaction is always relevant\cite{vidal,kh_h} for both diagonal and off-idagonal modulations. Even in the diagonal modulation case, the ground state becomes a Mott insulator if the Coulomb interaction is strong enough. In this case, however, the effective exchange modulation in the spin sector is irrelevant and spin sector behaves as an antiferromagnetic uniform Heisenberg chain. 

Our calculation suggests how to realize different types of Fibonacci electronic system using quantum dot arrays. To realize nearly free Fibonacci chain, the local potential of dots should be modulated and the charging energy should be reduced. On the other hand, to realize Fibonacci Heisenberg antiferromagnet, it is the transfer integrals (or distances) between the dots which should be modulated.

The numerical calculations were performed using the HITAC SR8000 at the Supercomputer Center, Institute for Solid State Physics, University of Tokyo and  the HITAC S820/80 at the Information Processing Center of Saitama University.  This work is supported by a Grant-in-Aid for Scientific Research from the Ministry of Education, Science, Sports and Culture, Japan.

\end{document}